\documentclass[eqsecnum,showpacs,twocolumn,pra,aps]{revtex4-1}

\usepackage{graphicx,amssymb,bm}
\usepackage{color}

\def\ket#1{|\,#1\,\rangle}
\def\bra#1{\langle\, #1\,|}

\def\proj#1#2{\ket{#1}\bra{#2}}
\def\expect#1{\langle\, #1\, \rangle}

\def\ol#1{\overline{#1}}

\def\ol#1{\overline{#1}}

\newcommand{\beq}{\begin{equation}}
\newcommand{\eeq}{\end{equation}}
\newcommand{\beqa}{\begin{eqnarray}}
\newcommand{\eeqa}{\end{eqnarray}}

\def\opone{\leavevmode\hbox{\small1\kern-3.8pt\normalsize1}}
\newcommand{\pgen}{\mathbb{P}}
\newcommand{\phs}{\mathbb{P}_{\rm HS}}
\newcommand{\pb}{\mathbb{P}_{\rm B}}
\newcommand{\ps}{\mathbb{P}_{{\rm RE}}}

\newcommand{\nh}{\hat N_H}
\newcommand{\nv}{\hat N_V}

\newcommand{\rop}{\hat{\rho}}
\newcommand{\sop}{\hat{\sigma}}

\begin{document}
\title{Modification of polarization through  de-Gaussification}

\author{Iulia Ghiu}
\email{iulia.ghiu@g.unibuc.ro}
\author{Paulina Marian}
\email{paulina.marian@g.unibuc.ro}
\author{Tudor A. Marian}
\email{tudor.marian@g.unibuc.ro}
\affiliation{Centre for Advanced  Quantum Physics,\\
Department of Physics, University of Bucharest, \\
R-077125 M\u{a}gurele, Romania}

\date{\today}

\begin{abstract}

We analyze the polarization of a quantum radiation field under a de-Gaussification process. Specifically, we consider the addition of photons to a two-mode thermal state to get mixed non-Gaussian and nonclassical states which are still diagonal in the Fock basis. Stokes-operator-based degrees of polarization and two distance-type measures defined with Hilbert-Schmidt and Bures metrics are investigated. For a better insight we here introduce a polarization degree based on the relative entropy. Polarization of the thermal states is fully investigated and simple closed expressions are found for all the defined degrees. The evaluated degrees for photon-added states are then compared to the corresponding ones for the two-mode thermal states they originate from. We present interesting findings which tell us that some popular degrees of polarization are not fully consistent. However, the most solidly defined degrees, which are based on the Bures metric and relative entropy, clearly indicate an enhancement of polarization through de-Gaussification. This conclusion is supported by the behavior of the degrees of polarization of the Fock states, which are finally discussed as a limit case.

\end{abstract}


\maketitle

\section{Introduction}

An important concept for classical optics, polarization of light has quite recently become of interest  in quantum information processing as well.  Its usefulness in this area  arises especially from the robustness of light as an information carrier. This  allows easy manipulation and transmission
of polarization-encoded information
with negligible
losses, thus providing an appreciated experimental convenience.
Indeed,  polarization encoding was considered to be  the optimal choice in many recent experiments: the quantum key
distribution required in cryptography \cite{Bennett-1992}, \cite{Muller}, polarization entanglement
\cite{Kwiat-1995},
superdense coding \cite{Bennett-prl-1992}, quantum teleportation of polarization \cite{Bouw}, entanglement
swapping \cite{Bose}, quantum tomography \cite{Barbieri},and quantum computation \cite{Joo}. Due to the relevance
 of polarization in such a large number of quantum processes, one needs to find proper measures for description
 of polarization in the quantum realm.

Classically, the definition of the degree of polarization is obtained by using  the Stokes parameters
\cite{Stokes,Mandel}. In the quantum domain, the standard degree of polarization was defined by
replacing the Stokes variables by the expectation values of the Stokes operators \cite{Fano-49,Collet-70,Simon,Kl-92}. This definition based on first-order moments cannot give a
complete description for all quantum fields, since it assigns the zero value in some cases of pure polarized
states. Therefore, the idea of construction of the polarization measures by using second-order moments
of the Stokes operators occured \cite{Alod,Klimov-2010}. Moreover, a provisionally improved characterization of
polarization has recently been obtained with the help of higher-order moments \cite{Bjork-2012,Hoz-2013,Sanchez-2013,Hoz-2014,Bjork-2015,Bjork-phys-scr-2015}. Collaterally, it was recently proved that the Stokes-operator measurements have  great importance for the estimation of the covariance
matrix of macroscopic quantum states \cite{Filip-2016}.

Taking inspiration from the quantum information tool-box, the degree of polarization was
quantified as the  distance between the given quantum field state and the set of all
unpolarized states. These definitions have considered several metrics: Hilbert-Schmidt, Bures
\cite{Klimov-2005,Bjork-2006} and Chernoff \cite{Ghiu-2010,Bjork-2010, Ghiu-rom-ac-2015,Ghiu-rom-j-2016,Ghiu-rom-rep-2018}.

Two recent reviews by
Chirkin  and Luis present possible definitions of the degree of polarization of a quantum field and applications of the polarized states \cite{Chirkin,Luis-2016}.  An
overview of the difficulties encountered in using various types of polarization degrees
was recently given in Ref. \cite{GJ}.

A lot of attention has been devoted in recent years to the study of polarization of Gaussian states of 
light fields \cite{Korolkova,Bowen,Luis-2007,Muller-2016}.  On the other hand, one would find little
investigation of polarization for non-Gaussian field states.  See, however, the interesting findings on the polarization of pure  Schr\"{o}dinger cat or cat-like states and entangled bimodal coherent states in Ref. \cite{Singh}.
Since the non-Gaussian states were proven to be more efficient  resources in some quantum information processes \cite{Opa,Olivares,Anno,Eisert,Cerf}, we felt the need to analyze  their polarization more deeply using some degrees originating from both the classical and the quantum perspectives.  Our aim here is thus twofold. First we are interested in observing the behavior of polarization under de-Gaussification as an interesting process in its own right. Second, we want to figure out the consistency of  the results given by differently defined degrees of polarization and eventually draw a conclusion on their usefulness.
Specifically, we  use a description of the polarization for
a product of mixed states which are Fock-diagonal. With the unique exception of the thermal states, these are definitely non-Gaussian.

The paper is organized as follows. In Sec. II we review the traditional Stokes definitions based
on the first- and second-order moments.  We further consider
two distance-type measures based on the Hilbert-Schmidt and Bures metrics.
In Sec. III, we introduce a new definition of the quantum degree of polarization which is based on the relative entropy.
The above-mentioned quantum
degrees of polarization are applied to a tensor product of Fock-diagonal states in Sec. IV. The obtained closed expressions are state-dependent expansions on   $N$-photon  manifolds.  In Sec. V we give our  exact findings on the degrees of polarization of two-mode thermal states. These are important results because we then compare them to the degrees of polarization of some non-Gaussian states resulting from the  addition of photons to  thermal states in Sec. VI. Our conclusions are drawn  in Sec. VII.

\section{Quantum degrees of polarization}

In any discussion of the polarization of the quantum radiation
field, a quasi-monochromatic light beam is decomposed into
two orthogonal transverse oscillating modes which are described
by a definite two-mode state $\hat \rho$. The quantum treatment
of polarization starts from the Stokes operators built with the amplitude operators of the conventional horizontally $(H)$ and vertically $(V)$ oscillating modes:
\begin{eqnarray*}
\hat S_1:=\hat a _{H}^\dagger \hat a_{V}+\hat a _{H} \hat a_{V}^\dagger ,
\qquad \hat S_2:=\frac{1}{i} \left( \hat a _{H}^\dagger \hat a_{V}
-\hat a _{H} \hat a_{V}^\dagger \right)  \end{eqnarray*}
\begin{eqnarray}\hat S_3:=
\hat a _{H}^\dagger \hat a_{H}-\hat a _{V}^\dagger \hat a_{V};\;\hat S_0:=\hat a _{H}^\dagger \hat a_{H}
+\hat a _{V}^\dagger \hat a_{V}. \label{St1}
\end{eqnarray}
Their expectation values correspond to the classical Stokes parameters:
\begin{eqnarray}\langle \hat S_0 \rangle={\rm Tr}(\hat \rho \hat S_0),\;
\langle \hat S_j \rangle={\rm Tr}(\hat \rho \hat S_j),
\; (j=1, 2, 3).\label{St2} \end{eqnarray}

\subsection{Previously defined measures}

The first proposal for defining a quantum degree of polarization based on Stokes variables was a
direct generalization of the classical measure \cite{Fano-49,Collet-70,Simon,Kl-92}:
\beq
\pgen_1({\rop }):=
\frac{\sqrt{\expect{\hat S_1}^2+\expect{\hat S_2}^2+\expect{\hat S_3}^2}}{\expect{\hat S_0}}.
\label{def-p-stokes}
\eeq
The index 1 emphasizes that this definition considers only first-order moments of the Stokes operators.
Accordingly, all the product-states $\ket{\psi}_H\ket{0}_V$ have a degree of
polarization equal to unity. For the two-mode state close to the vacuum, i.e. $\ket{\psi}_H\to \ket{0}_H$,
one obtains also $\pgen_1=1$, which is an unphysical  result \cite{Klimov-2010}.

Since $\pgen_1$ cannot be regarded as a proper definition of the quantum degree of polarization, proposals based on
higher-order moments have to be considered. A second-order quantum degree of polarization was introduced
in Ref. \cite{Alod}:
\beq
\pgen_2({\rop }):= \sqrt{1-\frac{(\Delta {\hat \bold S})^2}{\expect{\hat{\bold S}^2}}}.
\label{def-p2prim}
\eeq
Here $\hat \bold S$ is the Stokes vector whose components are written in Eqs.\ (\ref{St1}) and
$$(\Delta{\hat{\bold S}})^2:=(\Delta{\hat{ S_1}})^2 + (\Delta{\hat{ S_2}})^2+(\Delta{\hat{ S_3}})^2 $$ is the total variance
of the Stokes operators. We  get  further
\beq
\pgen_2({\rop })=
\sqrt{\frac{\expect{\hat S_1}^2+\expect{\hat S_2}^2+\expect{\hat S_3}^2}{\expect{\hat \bold S^2}}}.
\label{def-p2}
\eeq
This definition gives the correct answer for the state close to the vacuum:
$\pgen_2(\ket{\psi}_H \ket{0}_V)\to 0$ when $\ket{\psi}_H\to \ket{0}_H$.

 More recently, various distance-type degrees of polarization have also been investigated \cite{Klimov-2005,Bjork-2010}. Recall that in general, the distance of a given state having a specific property to a reference set of states not having it has been recognized as a measure of that property.
   The  essence of defining a reliable distance-type measure consists in choosing a convenient metric and identifying a appropriate reference set of states. Application of this recipe to the polarization issues  is greatly facilitated by  precise
   knowledge of the set ${\cal U}$ of unpolarized states.
 Indeed,  any unpolarized two-mode state $\hat \sigma$ has a block-diagonal sector $\hat \sigma_b$ \cite{Ghiu-2010}
which is  SU(2)-invariant
\cite{Prakash,Bjork5}:
\begin{equation}
\hat \sigma_b =\sum_{N=0}^\infty \; \pi_N \frac{1}{N+1}\, {\hat P}_N. \label{tau}
\end{equation}
In Eq.\  (\ref{tau}),
\begin{equation}
{\hat P}_N:=\sum_{n=0}^N |n,N-n \rangle \langle n,N-n| \label{projection}
\end{equation}
is the projection operator onto the vector subspace of the $N$-photon states,
called the $N$th excitation manifold. We have denoted
$|n,N-n \rangle:={|n \rangle}_H\otimes {|N-n \rangle}_V$.
Further, $\pi_N$ are the photon-number probabilities in the $SU(2)$-invariant
state $\hat \sigma_b$ and they satisfy the normalization condition
\begin{equation}
\sum_{N=0}^{\infty}\pi_N=1. \label{certainty}
\end{equation}
Note that  any SU(2)-invariant state\ (\ref{tau}) is Fock-diagonal and, except for the vacuum,  is mixed.

We recall two distance-type measures for the quantum degree of polarization based on Hilbert-Schmidt
and Bures metrics,which were defined in Ref. \cite{Klimov-2005}:
\beqa
\phs (\rop ) &=& \inf_{\sop \in {\cal U}}  \mbox{Tr} [(
\rop - \sop)^2 ], \label{polhs}  \\
\pb (\rop ) &=& 1 - \sup_{\sop \in {\cal U}} \sqrt{{\cal F} (\rop ,\sop )},
\label{polb}
\eeqa
where ${\cal U}$ represents the set of all unpolarized two-mode states, while ${\cal F}$ stands for the
fidelity between two states \cite{Jozsa},
\beq
{\cal F}({\hat \rho}_1,\,{\hat \rho}_2):=\left[\mbox{Tr}
\sqrt{{\hat \rho}_1^{1/2}{\hat \rho}_2\,{\hat \rho}_1^{1/2}}\right]^2.
\eeq

In Refs. \cite{Bjork-2010,Ghiu-2010} a quite different approach to defining quantum degrees of polarization  was proposed.
 The polarization properties of the given state ${\hat \rho}$ were delegated to its block-diagonal sector,
\begin{equation}
 {\hat \rho}_b:=
\sum_{N=0}^\infty {\hat P}_N{\hat \rho}{\hat P}_N \label{channel}.
\end{equation}
The polarization-relevant state\ (\ref{channel}) is the result of an ideal nonselective measurement of the observable $\hat N:=\hat N_{H}+\hat N_{V}$   which preserves the photon-number distribution of the given two-mode state
 ${\hat \rho}$  \cite{Ghiu-2010}.
 In particular, the quantum degree of polarization is defined as 
\beq \mathbb{P}(\hat \rho):= \mathbb{P}(\hat \rho_b), \label{op-dens-not} \eeq
  so that in Eqs.\ (\ref{polhs}) and\ (\ref{polb})  one should replace $\hat \rho$ with
$\hat \rho_b$  and $\hat \sigma$ with $\hat \sigma_b$.

Let us denote $\mu_{N,n}, \; (n=0,1,\cdots, N), $ the eigenvalues of the positive operator $ \hat P_N\hat \rho \hat P_N$.
Their sum is precisely the probability of the $N$th excitation manifold:
\beq
p_N={\rm Tr} (\hat \rho \hat P_N) =\sum_{n=0}^N\mu_{N,n}
\label{pn}
\eeq
The $N$-photon state $\hat \rho_N:=\frac{1}{p_N} \hat P_N\hat \rho \hat P_N,\; (p_N> 0)$,  commutes with any SU(2)-invariant state $\hat \sigma_b$, Eq.\ (\ref{tau}), and so does the polarization state $\hat \rho_b$:
     
\begin{equation}
[{\hat \rho}_b,{\hat \sigma}_b]=\hat 0. \label{com}
\end{equation}
Extremization of  expressions (\ref{polhs}) and
(\ref{polb}) of the Hilbert-Schmidt and Bures measures was previously carried out  \cite{Klimov-2005,Bjork-2010,Ghiu-2010}  by applying the method of  Lagrange multipliers.  An important and helpful property for the ongoing evaluations
is the commutation relation\ (\ref{com}). We write here the following general 
expressions in terms of the photon-number probabilities $p_N$ and the eigenvalues 
$\mu_{N,n}$:
\beqa
\phs (\hat \rho ) &=&\sum_{N=0}^\infty \sum_{n=0}^N \mu_{N,n}^2
-\sum_{N=0}^\infty \frac{p_N^2}{N+1},\label{hs} \\
\pb (\hat \rho ) &=&1-\sqrt{\sum_{N=0}^\infty \frac{1}{N+1}\,
\bigg( \sum_{n=0}^N\sqrt{\mu_{N,n}}\bigg) ^2}.\label{bures}
\eeqa
An interesting case to examine  is the polarization of an arbitrary pure state conveniently written as an expansion in pure $N$-photon states
\beq |\Psi\rangle=\sum_{N=0}^{\infty} c_N |\Psi_N\rangle. \label{pure1} \eeq
We have $p_N=|c_N|^2$ and $\sum_{N=0}^{\infty} p_N=1.$
Accordingly, the associate state ${\hat \rho}_b $ is the mixture
\beq {\hat \rho}_b =\sum_{N=0}^{\infty} |c_N|^2  |\Psi_N\rangle\langle \Psi_N |. \label{pure2} \eeq
Equations\ (\ref{hs}) and\ (\ref{bures}) greatly simplify to
\beqa
\phs ( |\Psi \rangle\langle \Psi |) &=&\sum_{N=0}^\infty |c_N|^4
\frac{N}{N+1},\label{hsp} \\
\pb ( |\Psi \rangle\langle \Psi |) &=&1-\sqrt{\sum_{N=0}^\infty \frac{|c_N|^2}{N+1}}.\label{buresp}
\eeqa

\section{Relative entropy as a measure of quantum polarization}
Despite not being a true metric, the relative entropy is acceptable as a measure of polarization due to its outstanding  distinguishability properties as discussed in Ref. \cite{Ved}.
Recall that the relative entropy between  state ${\hat \sigma}^{\prime}$
and  state ${\hat \sigma}^{\prime \prime}$ is defined as the difference \cite{Weh}
\begin{equation}
{\cal S}(\hat \sigma^{\prime }|\hat \sigma^{\prime \prime}):={\rm Tr} [\hat \sigma^{\prime }\ln(\hat \sigma^{\prime})]-
{\rm Tr} [\hat \sigma^{\prime }\ln(\hat \sigma^{\prime \prime})].
\label{re}
\end{equation}
The relative entropy was successfully used as a measure of entanglement
for pure bipartite states providing one of the few exact and general evaluations \cite{VP}. A more recent  general result \cite{PT2013} finds the relative entropy to be an exact measure of  non-Gaussianity.

In view of the preceding discussion
on  the appropriateness of the associate  block-diagonal density operator ${\hat \rho}_b$ in describing polarization, we define a degree of polarization based on the relative entropy as
\beq
\ps (\rop ):=\inf_{\sop \in {\cal U}} \, \frac{S(\rop_b |\sop_b)}{1+S(\rop_b |\sop_b)}.
\label{pol-entrop-def}
\eeq
 The relative entropy between the commuting states $\hat\rho_b$ and  $\hat\sigma_b$, Eq.\ (\ref{tau}), is 
 \beq
S( \rop_b|\hat{\sigma}_b)=-S(\rop_b)-\sum_{N=0}^\infty p_N\, \ln \left(\frac{\pi_N}{N+1}\right).
\label{entr-rel}
\eeq
In Eq.\ (\ref{entr-rel}),  $S(\rop)=- {\rm Tr} [\hat \rho \ln (\hat \rho)]$ is the von Neumann entropy of state $\hat\rho$ and  $p_N$ is given by Eq. (\ref{pn}).
Our task is to evaluate the parameters $\tilde\pi_N$ of the unpolarized state   ${\hat{\tilde \sigma}_b}$ for which the infimum in Eq.\  (\ref{pol-entrop-def}) is realized. As far as we know at this moment, the present work is the first  to look for the closest unpolarized two-mode state
through relative entropy.
We have to minimize  the function\ (\ref{entr-rel}) with respect to the probabilities $\pi_N$
under the constraint\ (\ref{certainty}).  Similarly to  what was previously discussed regarding other distance-type measures \cite{Klimov-2005,Bjork-2010,Ghiu-2010}, the extremization is easily performed by applying the method of Lagrange multipliers.
We easily get the conditions of minimum
\beq \tilde \pi_N=p_N, \;(N=0,1,2,3,\dots). \eeq
Interestingly, the closest unpolarized state is the same as for the Hilbert-Schmidt polarization measure first written in Ref.\cite{Klimov-2005}. It has the same photon-number distribution as the given state,
\beq {\hat{\tilde\sigma}_b}=\sum_{N=0}^\infty \; p_N \frac{1}{N+1}\, {\hat P}_N, \label{tau1}
\eeq
and leads to the final formula
\beq
S( \rop_b|{\hat{\tilde\sigma}_b})=-S(\rop_b)-\sum_{N=0}^\infty p_N\, \ln \left(\frac{p_N}{N+1}\right).
\label{entr-rel2}
\eeq
For the pure state\ (\ref{pure1}),  the minimal relative entropy\ (\ref{entr-rel2}) simplifies to
\beq
S( \rop_b|{\hat{\tilde\sigma}_b})=\sum_{N=0}^\infty |c_N|^2 \ln (N+1).\label{pureRE}\eeq

\section{Polarization of  Fock-diagonal states }
Obviously, any Fock-diagonal state $\hat \rho$ coincides with its polarization sector, namely, 
$\hat \rho_b= \hat \rho$. 
For the sake of simplicity, in this paper we deal with a mixed Fock-diagonal product state,
\beqa
\hat \rho = \hat \rho_H\otimes \hat \rho_V, \label{rho}
\eeqa
where
\beqa
\hat \rho_H=\sum_{m=0}^\infty \xi_m\, \proj{m}{m}, \;\;
\hat \rho_V=\sum_{n=0}^\infty \eta_n\, \proj{n}{n}. \label{states}
\eeqa
The photon-number distributions  $\xi_m$ and $\eta_m$  satisfy the normalization conditions
$$\sum_{m=0}^\infty \xi_m =1,\;\;\; \sum_{m=0}^\infty \eta_m =1.$$

Our first aim here is to write  the quantum degrees of polarization reviewed
in the previous section for this type of Fock-diagonal states. With the notable exception
of the thermal states, we  thus deal with non-Gaussian density operators which are known to be important in several protocols of quantum information.
To begin,  let us recall  the photon-number operators in the two modes,
i.e. $\nh = \hat{a}^\dagger_H \hat{a}_H $ and $\nv =
\hat{a}^\dagger_V \hat{a}_V$. The expectation values of the Stokes operators\ (\ref{St1}) are found to be
$$\expect{\hat S_1}=\expect{\hat S_2}=0, \expect{\hat S_3}=\expect{\nh}-\expect{\nv}$$
and further
$$\expect{\hat{\bold S}^2}=
2 (\expect{\nh }\expect{\nv} +\expect{\nh }+\expect{\nv })+ \expect{\nh ^2} +\expect{\nv^2}.$$
In general we simply find
\begin{eqnarray*}
\expect{\nh^j}=\sum_{m=0}^\infty \xi_m\, m^j, \; \; \expect{\nv^j }=\sum_{m=0}^\infty \eta_m\, m^j,\;\; (j=1,2,\cdots).
\end{eqnarray*}

 By using Eq. (\ref{def-p-stokes}) we get the first-order Stokes degree of polarization
\beqa
\pgen_1(\hat \rho)= \frac{\vert \expect{\nh }-\expect{\nv }\vert}{\expect{\nh }+
\expect{\nv } }.
\label{p-stokes}
\eeqa

Further, Eq. (\ref{def-p2prim}) becomes in this case
\beq
\pgen_2(\hat \rho)= \frac{\vert \expect{\nh }
-\expect{\nv }\vert}{\sqrt{\expect{\hat{\bold S}^2}}}.
\label{p2}
\eeq
As regards the distance-type measures, they are simply written by setting $\mu_{N,n}=\xi_n \eta_{N-n}$  into  Eqs.(2.16), (2.17) and (3.6).

As an example, the entropic degree of polarization is given by Eq.\ (\ref{pol-entrop-def}) after inserting
 the explicit relative entropy of polarization for the state\ (\ref{rho})
\beq
S(\rop |\hat{\tilde \sigma}_b)= \sum_{m=0}^{\infty} [\xi_m\ln (\xi_m)+\eta_m\ln(\eta_m)] -\sum_{N=0}^\infty p_N \ln \left(\frac{p_N}{N+1}\right).\label{pol-entrop}
\eeq
In the following we specialize
 the above final expressions of the quantum degrees of polarization  to two interesting Fock-diagonal states: the unique Gaussian case which is a two-mode thermal state and a non-Gaussian example prepared by adding photons to a thermal state.

\section{Polarization of two-mode thermal states}

We here compute the quantum degrees of polarization for the relevant class of
two-mode thermal states whose  density operators depend only on the mean occupancies $\bar n_1:=\langle \hat N_H \rangle,\bar n_2:=\langle \hat N_V \rangle$ of the modes:
\beq   \rop_{\rm{th}}(\bar n_1, \bar n_2): = \rop_{\rm{th}}(\ol n_1)\otimes
\rop_{\rm{th}}(\ol n_2)\label{st-term}
\eeq with
$$
\rop_{\rm{th}}(\ol n_j)=\frac{1}{\ol{n}_j+1}\sum_{n=0}^\infty
\left( \frac{\ol{n}_j}{\ol{n}_j+1}\right)^n\ket{n}\bra{n}, (j=1,2).
$$
It is convenient to rewrite the  thermal state\ (\ref{st-term})
as follows:
\begin{eqnarray}   \rop_{\rm{th}}(\bar n_1, \bar n_2)& =&(1-q_1)(1-q_2)  \nonumber \\&& \sum_{N=0}^{\infty}\sum_{n=0}^{N} q_1^n q_2^{N-n} \proj{n,N-n}{n,N-n}, \nonumber \\
\label{ST1}\end{eqnarray}
where
 the notation $q_j:={\ol{n}_j}/{(\ol{n}_j+1)}$ has been introduced.
Before proceeding with the evaluation of various polarization degrees we have to notice an important property.
At thermal equilibrium ($q_1=q_2:=q$) the density operator\ (\ref{ST1}) simplifies to
\beq
\rop_{\rm{th}}(\bar n,\bar n)=(1-q) ^2
\sum_{N=0}^\infty q^N \hat P_N
\label{NST}
\eeq
According to Eq.\ (\ref{tau}), the thermal state at equilibrium, Eq.\ (\ref{NST}), is unpolarized   and
therefore its degree of polarization should  be equal to zero, regardless of the type of measure we use.
In the following we take the thermal equilibrium limit
as an useful test of our results.

\subsection{Degrees of polarization based on the Stokes operators}

We insert the thermal mean occupancies and the expectation values
$$\langle \hat N_H^2 \rangle=\bar n_1(2 \bar n_1+1),\;\;\langle \hat N_V^2 \rangle=\bar n_2(2 \bar n_2+1)$$
in Eqs.(\ref{p-stokes}) and (\ref{p2})  to easily get
\beq
\pgen_1(\rop_{\rm{th}}(\bar n_1, \bar n_2))=
\frac{\vert \ol n_1-\ol n_2\vert}{\ol n_1+\ol n_2}.
\eeq

\beq
\pgen_2(\rop_{\rm{th}}(\bar n_1, \bar n_2))=\frac{\vert \ol n_1-\ol n_2\vert}{\sqrt{2\, \ol n_1^2+2\, \ol n_2^2+2\, \ol n_1\, \ol n_2+3\, \ol n_1+3\,
\ol n_2}}.
\eeq
Both expressions were first written in Ref.\cite{Luis-2007}. They obviously meet the requirement of being equal to 0 at thermal equilibrium.

\subsection{Distance-type degrees of polarization}

Two distance-type measures of  polarization can be evaluated  analytically for a thermal state, Eq.\ (\ref{ST1}), due to the privilege of getting a closed form for the probability of an $N$-photon state:
\begin{eqnarray}
p_N=(1-q_1)(1-q_2) \sum_{n=0}^N q_1^n q_2^{N-n}\nonumber \\
=(1-q_1)(1-q_2)\frac{q_1^{N+1}-q_2^{N+1}}{q_1-q_2}.\label{TS2}
\end{eqnarray}
Above we have simply used the geometric sequence
\begin{eqnarray*}
\sum_{n=0}^{N}q^n=\frac{1-q^{N+1}}{1-q}.\end{eqnarray*}
For $0\leqq q<1$ and $N\rightarrow \infty$, we get the well known geometric series
\begin{eqnarray}
\sum_{n=0}^{\infty}q^n=\frac{1}{1-q}, \;\; (0\leqq q<1),\label{GS}\end{eqnarray}
whose  term-by-term integration  yields another useful power series
 \begin{eqnarray}
\sum_{n=0}^{\infty}\frac{1}{n+1}q^{n+1}=-\ln (1-q).\;\;(0\leqq q<1).\label{GS1}\end{eqnarray}
Evaluation of the Hilbert-Schmidt polarization degree, Eq.(\ref{hs}), is routinely performed via Eqs.\ (\ref{TS2})
and\ (\ref{GS1}). In terms of thermal mean occupancies we nicely get
\beqa
\phs (\rop_{\rm{th}}(\bar n_1, \bar n_2)) &=&\frac{1}{\left( 2\ol{n}_1+1\right)
\left( 2\ol{n}_2+1\right)} \nonumber \\
&-&\frac{1}{\left( \ol{n}_1-\ol{n}_2\right)^2}\ln \left[ \frac{\left( \ol{n}_1+
\ol{n}_2+1\right)^2}{\left( 2\ol{n}_1+1\right)\left( 2\ol{n}_2+1\right)}\right].\nonumber\\
\label{hs1}
\eeqa
In the symmetric case $\bar n_1=\bar n_2$  the obtained degree goes to 0 as it should. However, Eq.\ (\ref{hs1})
reveals a non-monotonic behavior of the Hilbert-Schmidt degree which has the limit zero when the difference between the two thermal mean occupancies is very large. This suggests that the Hilbert-Schmidt metric is not a reliable measure of polarization.

According to Eq.(\ref{bures}), in order to obtain the Bures degree of polarization for a two-mode
thermal state we need to use once more the geometric sequence to write
$$
\sum_{n=0}^N\sqrt{\mu_{N,n}}=[(1-q_1)(1-q_2)]^{1/2}\; \frac{q_2^{\frac{N+1}{2}}-
q_1^{\frac{N+1}{2}}}{{q_2^{1/2}-q_1^{1/2}}}.
$$
By replacing the above result in Eq.(\ref{bures}) and again taking advantage of the series\ (\ref{GS1}) we write  the Bures degree of polarization:
\beq
\pb (\rop_{\rm{th}}(\bar n_1, \bar n_2))=1-\frac{ \sqrt{2\; \ln
\left[ \sqrt{\left( \ol{n}_1+1\right) \left( \ol{n}_2+1\right)}-\sqrt{\ol{n}_1\ol{n}_2}\right]}   }{| \sqrt{\ol{n}_1
\left( \ol{n}_2+1\right)}-\sqrt{\ol{n}_2\left(\ol{n}_1+1\right)}|}.\label{B1}
\eeq
The Bures degree\ (\ref{B1}) appears to be monotonic and is 0 in the symmetric case.

The last degree we have to examine is the entropic one, Eq.\ (\ref{pol-entrop}). With
the von Neumann entropy of a one-mode thermal state,
\beq
S(\rop_{\rm{th}}(\ol n))=(\ol n+1)\, \ln (\ol n+1)-\ol n\, \ln (\ol n),\label{VNE}
\eeq
and the explicit expression of the probability of an $N$-photon state, Eq.\ (\ref{TS2}), we are  left with the numerical evaluation of the sum appearing in the expression of the minimal relative entropy:
\beq 
S(\rop|\hat{\tilde \sigma}_b)=-S(\rop_{\rm{th}}(\ol n_1))-S(\rop_{\rm{th}}(\ol n_2))\nonumber \\-\sum_{N=0}^\infty p_N\, \ln \left(\frac{p_N}{N+1}\right).
\eeq

\begin{figure}
\centering
\includegraphics[width=8cm]{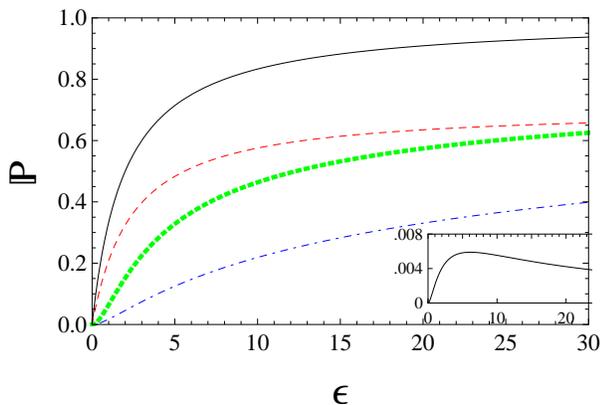}
\caption{Different degrees of polarization of the two-mode thermal state in terms of $\epsilon =|\ol n_1-\ol n_2|$:  Stokes degrees $\pgen _1$ (solid black curve) and $\pgen _2$ (dashed red curve), relative entropy (green, dotted), Bures (blue, dot-dashed). The inset shows the non-monotonic behaviour of the Hilbert-Schmidt measure. We have considered $\ol{n}_2=1$.}
\label{st-termica-toate}
\end{figure}
Figure 1 shows a monotonic aspect  of either Stokes-operator-based degrees or Bures and entropic ones versus the relative mean occupancy $\epsilon: =|\ol n_1-\ol n_2|$  of the two modes.  These degrees are consistent having a similar
 behavior with respect to the same parameter. Quite the reverse, in the inset we see the Hilbert-Schmidt  polarization degree\ (\ref{hs1})
 displaying a flat maximum and then  slowly decreasing to 0.
 The results obtained in this section on the polarization of the thermal states are compared in the following to the similar degrees for a
class of non-Gaussian ones prepared by adding photons to thermal states.

\section{Polarization of non-Gaussian states}

 It is well known that various
excitations on a single-mode thermal state of the type $\hat \rho \sim (\hat a^{\dag})^k \hat a^l \hat \rho_{\rm th}
(\hat a^{\dag})^l \hat a^k $  are diagonal in the Fock basis. Here $\hat a$ and $\hat a^{\dag}$ are the amplitude operators of the field mode. In general, states with added photons are non-classical and non-Gaussian.
We choose to apply now both Stokes-operator-based and distance-type degrees of polarization to a class of states of this type, that is, their density operator is the tensor product shown in Eqs.\ (\ref{rho})-(\ref{states}). Specifically,  we work here with a tensor product of multi-photon-added thermal states. Single-mode photon-added thermal states (PATSs) were introduced in Ref.\cite{AT} where  their non-classicality  expressed by the negativity of the $P$ function was studied. Quite recently, thermal states  with one-added photon were experimentally prepared and their non-classical properties were investigated
 \cite{bel,bel1,Kis}.  The present authors  were interested in the non-classicality of PATSs  and looked at  their evolution  during the interaction of the field mode with a thermal reservoir. Basically we have  investigated  two processes: loss of nonclassicality indicated by the time development of the Wigner and $P$ functions and loss of non-Gaussianity shown by some recently introduced distance-type measures \cite{Marian-2013,Marian-ps-2013}.

  Adding $M$ photons to a thermal state $\rop _{\rm{th}}(\ol n)$ results in the density operator
  \beq
\hat \rho^{(M)}(\ol n)=\frac{1}{M!\, (\bar n+1)^M}\, (\hat a^\dagger )^M\, \hat \rho_{\rm{th}}(\ol n)\,
\hat a^M,
\label{phadd}
\eeq
which in the Fock basis is easily written as  the mixture \cite{Marian-2013}:
\beq
\rop^{(M)}(\ol n)=\sum_{l=M}^\infty \left( \begin{array}{c}
l\\ M
\end{array}
\right)\frac{{\bar n}^{l-M}}{(\bar n+1)^{l+1}}\, \proj{l}{l}.
\label{ph-add}
\eeq
The purity of a PATS was found in Ref. \cite{Marian-ps-2013}
 as a function of the thermal ratio $q=\bar n/(\bar n+1)$, which  involves a Legendre polynomial:
\begin{eqnarray}   {\rm Tr}\left\{\left[{\hat \rho}^{(M)}(\bar n)\right]^2\right\}
=\left(\frac{1-q}{1+q}\right)^{M+1}
{\cal P}_M\left(\frac{1+q^2}{1-q^2}\right). \label{pl4}
\end{eqnarray}
Note that the above Legendre polynomial ${\cal P}_M$  is strictly positive because its argument is at least equal to 1.

For polarization issues we consider the tensor product of two PATSs
\beq \rop:= \rop^{(M)}(\ol n_1)\otimes \rop^{(S)}(\ol n_2).\label{TP}\eeq
\subsection{Degrees of polarization based on Stokes operators}
In order to evaluate the quantum degrees of polarization based on the Stokes variables we only need
 the expectation values of the operators $\hat N$ and
$\hat N^2$ of the PATS $\rop^{(M)}(\ol n)$.  These can be obtained with the photon-number distribution of PATSs  arising from Eq.\ (\ref{ph-add}). More elegantly, we can use  the generating function of PATSs written  in Ref.\cite{Marian-ps-2013} to get:
\beqa
\expect{\hat N}&=& M(\ol n+1)+\ol n; \nonumber \\
\expect{\hat N^2}&=& \ol n (M+1) [(M+2) \ol n+2M+1]+M^2.
\label{medii-st-phot-add}
\eeqa

\begin{figure}[h]
\centering
\includegraphics[width=7cm]{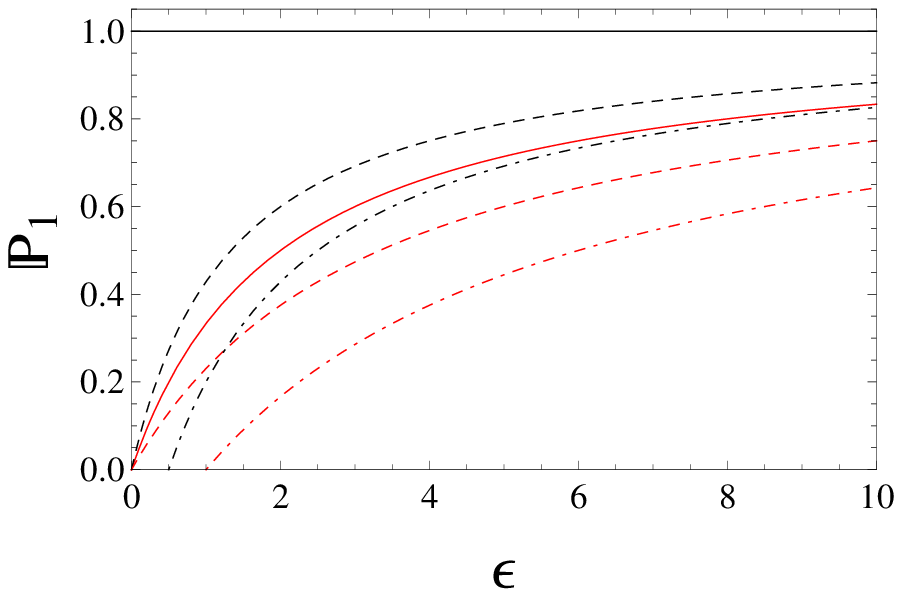}
\includegraphics[width=7cm]{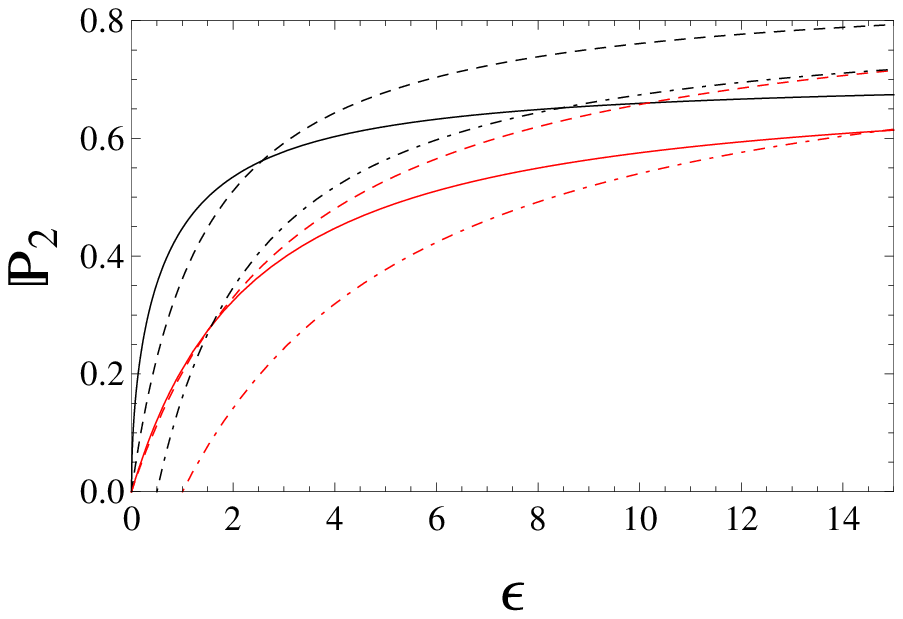}
\caption{Stokes degrees of polarization\ (\ref{def-p-stokes}) (top)  and\ (\ref{def-p2}) (bottom) in terms of $\epsilon =|\ol n_1-\ol n_2|$ for two classes of states: the two-mode thermal state (solid lines) and two-mode photon-added thermal state (dashed and dot-dashed lines). We have considered $\ol{n}_2=0$ (black curves) and $\ol{n}_2=1$ (red curves) and, in addition, two cases of the two-mode photon-added thermal state: the symmetric one $M=S=2$ (dashed curves) and the non-symmetric one $M=1$, $S=2$ (dot-dashed curves).}
\label{fig-comp-pol-1}
\end{figure}

In the following,  all  figures representing  various polarization degrees are plotted versus the relative thermal mean occupancy $\epsilon =|\ol n_1-\ol n_2|$  of the two modes. We used  the same values of the parameters of the PATSs in order to facilitate the comparison of their behaviors. Specifically, the black plots are characterized by $\bar n_2=0$. That is,  we deal with the particular state
\beq \rop:= \rop^{(M)}(\ol n_1)\otimes |S\rangle \langle S|,\label{TF}\eeq
which is a tensor product of a PATS and a Fock state.   The red plots describe the polarization of a proper two-mode
PATS, Eq.\ (\ref{TP}), for a fixed value $\bar n_2=1.$  For both sets of plots (black and red),  we examine  symmetric PATSs ($M=S=2$) and non-symmetric ones ($M=1, S=2$). Also plotted  (solid lines) in all subsequent figures are the same degrees of polarization for the  thermal states from which the corresponding PATSs are prepared.

Inserting Eqs. (\ref{medii-st-phot-add}) written for both $H$ and $V$  modes into the formulae (\ref{p-stokes}) and  (\ref{p2}), one obtains the expression of the two Stokes degrees of polarization described in Sec.II.
In Fig.\ref{fig-comp-pol-1} we plot the degrees $\pgen_1$ and $\pgen_2$ respectively under the conditions and parameters described above.
What can we see in these plots? The degree $\pgen_1$ based on only first-order moments of the Stokes operators indicates that by adding photons to a thermal state we are decreasing its polarization. This effect is stronger for non-symmetric addition. However $\pgen_1$  appears to be monotonic and consistent for either fixed values of $\bar n_2$.
Unlike  the aspect of $\pgen_1$, the degree  $\pgen_2$  displays an acute lack of consistency. The difference in  the polarization of thermal states and PATSs described by $\pgen_2$ fluctuates with their relative thermal mean occupancy.

\subsection{Distance-type degrees of polarization}

In order to calculate  the distance-type polarization degrees introduced in Sec.III for the state\ (\ref{TP})  we first write  the probability of its $N$th  excitation manifold, Eq.\ (\ref{pn}),
\beq p_N=\sum_{l=0}^N  \left( \begin{array}{c}
l\\ M
\end{array}
\right)  \left( \begin{array}{c}
N-l\\ S
\end{array}
\right) \frac{{\bar n_1}^{l-M}}{(\bar n_1+1)^{l+1}}\frac{{\bar n_2}^{N-l-S}}{(\bar n_2+1)^{N-l+1}}.
\label{pnMS}\eeq
This greatly simplifies for the particular state\ (\ref{TF}):
\beq p_N=  \left( \begin{array}{c}
N-S\\ M
\end{array}
\right) \frac{{\bar n_1}^{N-S-M}}{(\bar n_1+1)^{N-S+1}}.\label{pnTF}\eeq
For evaluating the Hilbert-Schmidt degree of polarization we use Eq. (\ref{hs}), as well as Eq. (\ref{pl4}) for the degree of purity of a PATS. The Hilbert-Schmidt measure in terms of thermal mean occupancies $\bar n_1$ and $\bar n_2$ is then
\begin{eqnarray*}
&&\phs \left( \rop^{(M)}(\ol n_1)\otimes \rop^{(S)}(\ol n_2) \right) =\frac{1}{(2\, \ol n_1+1)^{M+1}}\,
\frac{1}{(2\, \ol n_2+1)^{S+1}}\nonumber \\ &&\times{\cal P}_M \bigg(1+ \frac{2\, \ol n_1^2}{2\, \ol n_1+1} \bigg)\,
{\cal P}_S \bigg(1+ \frac{2\, \ol n_2^2}{2\, \ol n_2+1} \bigg)-\sum_{N=0}^\infty
\frac{p_N^2}{N+1},
\end{eqnarray*}
with  ${\cal P}_L(x)$ being the Legendre polynomial of degree $L$.

\begin{figure}
\centering
\includegraphics[width=7cm]{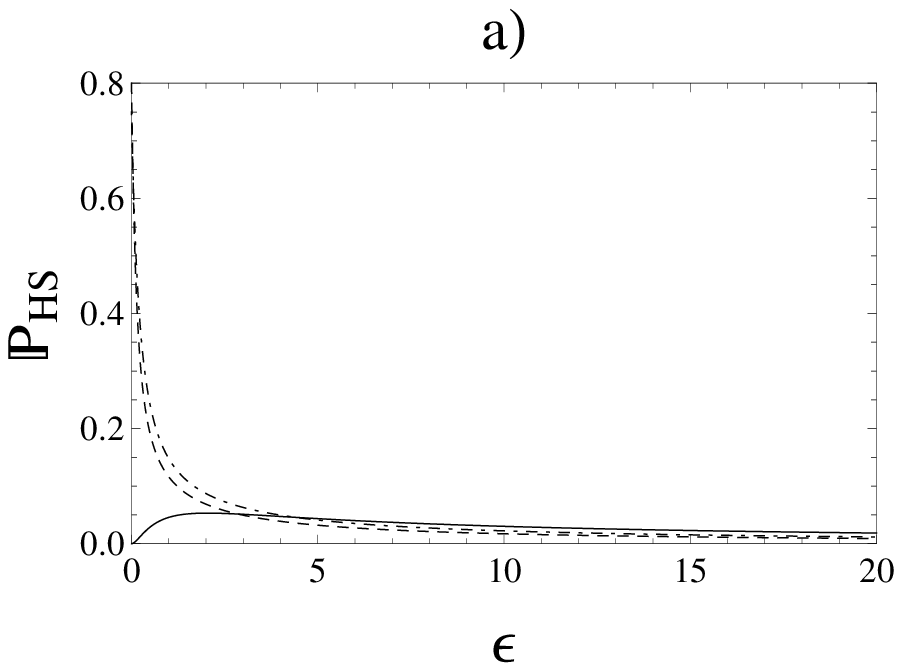}
\includegraphics[width=7cm]{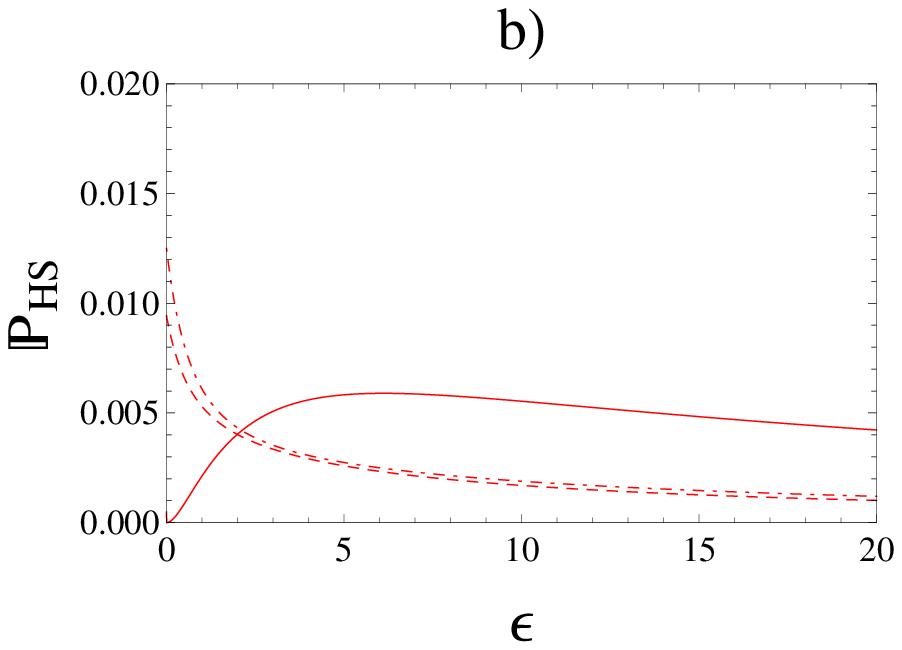}
\caption{Evaluation of the Hilbert-Schmidt degree of polarization in terms of $\epsilon =|\ol n_1-\ol n_2|$ for two classes of states: the two-mode thermal state (solid lines) and two-mode photon-added thermal state (dashed and dot-dashed lines). The parameters of two-mode PATSs are the same as in Fig.2.}
\label{fig-comp-hs}
\end{figure}

Figure \ref{fig-comp-hs} presents plots of
the Hilbert-Schmidt degree of polarization $\phs$ for  two-mode PATSs compared to the corresponding one for thermal states. We can see that adding photons to a thermal state strongly modifies the aspect of this  polarization degree in contrast with the evolution shown in Fig. 2 for Stokes-operator-based degrees.

Using the probability of its $N$th  excitation manifold written in Eqs.\ (\ref{pnMS}) and\ (\ref{pnTF}), we have numerically calculated the Bures measure, Eq.\ (\ref{bures}),  and  the entropic one, Eq.\ (\ref{pol-entrop}), for the two-mode thermal state and two-mode PATSs. To accomplish  this  task we have used
the von Neumann entropy of a PATS written in a simplified form:
\begin{eqnarray}
S(\hat \rho^{(M)}(\ol n))&=&(M+1) S(\rop_{\rm{th}}(\ol n))  \nonumber \\&&
-\sum_{l=M}^{\infty} \left(\hat \rho^{(M)}(\ol n)\right)_{ll}\ln \left[\left( \begin{array}{c}
l\\ M
\end{array}
\right)\right],
\end{eqnarray}
with $ S(\rop_{\rm{th}}(\ol n))$  being the von Neumann entropy of the thermal state, Eq.\ (\ref{VNE}).

\begin{figure}

\centering
\includegraphics[width=7cm]{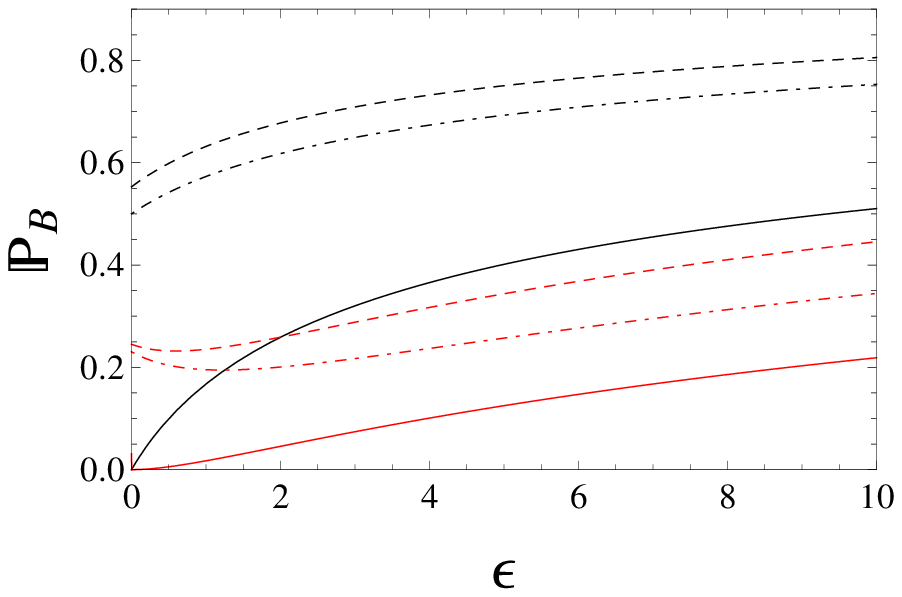}
\includegraphics[width=7cm]{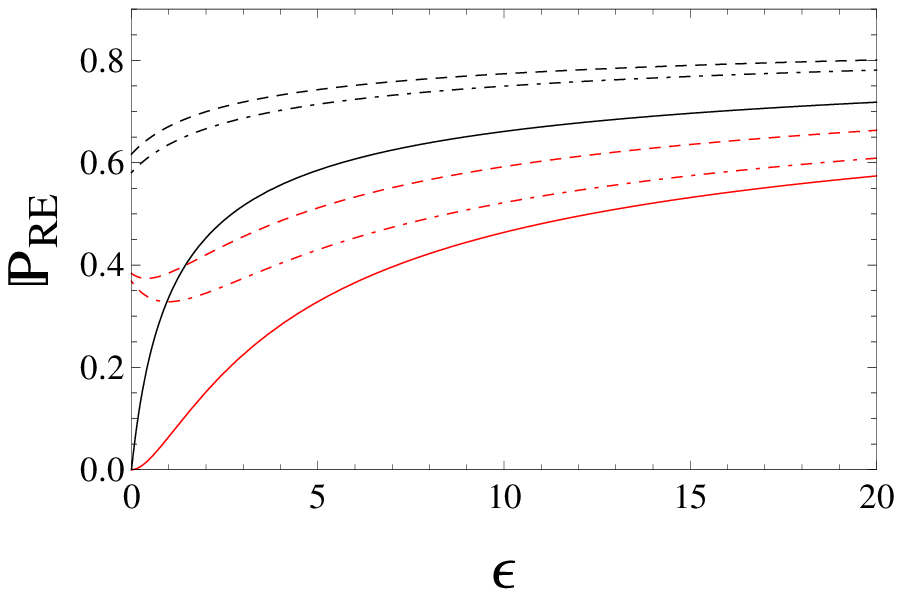}
\caption{ Bures (top) and relative entropy (bottom) degrees of polarization as defined in the text versus the relative thermal mean occupancy  $\epsilon =|\ol n_1-\ol n_2|$ for the same states and parameters  considered in Figs. 1-3. }
\label{fig-comp-bures}
\end{figure}

 For proper comparison of all these degrees, when plotting our final figure \ref{fig-comp-bures},  we have considered the same states and parameters as in Figs 2 and 3. What can we say about the plots in Fig. 4? They appear to be in agreement and manifest an overall consistency. Unlike the Hilbert-Schmidt degree, they show a monotonic behavior with respect to
the relative thermal mean occupancy.  Consistency means here the same ordering of the degrees for the same states in both cases.

\subsection{A limit case: polarization of a two-mode Fock state}
When setting $\ol n_1=0$ in Eq.\ (\ref{TF}), we get the density operator of a two-mode Fock state $|M,S\rangle$. The expressions of the degrees of polarization for this pure state with $N=M+S$ photons easily emerge as follows.
The series expansions on the right-hand side of Eqs.\ (\ref{hsp}),\ (\ref{buresp}) and\ (\ref{pureRE}) reduce to a single term:
\beq
\phs =
\frac{N}{N+1},\;
\pb =1-\sqrt{\frac{1}{N+1}},\;\ps=1-\frac{1}{1+ \ln (N+1)}.\label{pureN} \eeq
Note the obvious inequalities
\beq
\phs \geqq \pb;\;\; \phs \geqq \ps. \label{dtd}\eeq
While the only $N$-dependence is common to the above three distance-type degrees, the Stokes-operator-based degrees
depend on the difference between the occupancies of the two modes, just as  for classical light \cite{Mandel}. We get
\beqa
\pgen_1=&&
\frac{\vert M-S\vert}{N},\;\;
\pgen_2=\frac{\vert M-S\vert}{\sqrt{N(N+2)}},\nonumber\\
\pgen_1 \geqq \pgen_2. \label{stokesd}\eeqa
Comparing the above degrees to those for thermal light in Sec. V,  it appears that the Stokes-operator-based ones  are not quite sensitive to the statistical properties of the radiation. The distance-type degrees are all monotonic for the highly non-classical Fock states which is not the case for  thermal states  as shown in Fig. 1.

\section{Conclusions}

 We have examined the polarization of two-mode photon-added thermal states which are known to be non-classical, non-Gaussian and diagonal in the  Fock basis, in comparison to the thermal states from which they originate. The latter are  classical and the only Fock-diagonal Gaussian states.  Use was made of two types of degrees of polarization: one defined with Stokes operators by analogy with a classical description, the other being a bunch of distance-type  measures based on Hilbert-Schmidt and Bures metrics as well as on the relative entropy. We have first given a more general
 treatment valid for any Fock-diagonal states. Specializing it to the thermal states case, we have written for the first time closed and simple expressions for their Bures  and Hilbert-Schmidt degrees of polarization.

 Adding photons to thermal states is a deGaussification process which is currently being investigated in experiments. Modification of polarization due to this procedure is one of our interests in the present paper.  We have found that, according to the Stokes-operator-based degrees, polarization  decreases upon  adding photons to thermal states,  as shown in Fig.2. On the contrary, when looking at the Bures and entropic degrees in Fig.4,  polarization is larger for photon-added states.
 The Hilbert-Schmidt degree is not reliable due to its lack of consistency. So we are now in a dilemma regarding the description of polarization with the two types of defined degrees. How can we solve this? The agreement between the Bures degree and the entropic one is quite remarkable. However, the only solid conclusion that one can come to is that a supplementary testing of polarization is at this moment highly desirable. Work along these lines is in progress.

\section*{Acknowledgments}
 This work was supported by the funding agency CNCS-UEFISCDI of the Romanian Ministry of Research 
and Innovation through grant No. PN-III-P4-ID-PCE-2016-0794.


\begin{thebibliography}{99}
\bibitem{Bennett-1992} C. H. Bennett, F. Bessette, G. Brassard, L. Salvail, and J. Smolin, Experimental quantum cryptography,  J. Cryptology {\bf 5}, 3 (1992).

\bibitem{Muller} A. Muller, T. Hertzog, B. Huttner, W. Tittel, H. Zbinden, and N. Gisin, ``Plug and play'' systems for quantum cryptography, Appl. Phys. Lett. {\bf 70}, 793 (1997).

\bibitem{Kwiat-1995} P. G. Kwiat, K. Mattle, H. Weinfurter, A. Zeilinger, A. V. Sergienko, and Y. Shih,
New High-Intensity Source of Polarization-Entangled Photon Pairs, Phys. Rev. Lett. {\bf 75}, 4337 (1995).

\bibitem{Bennett-prl-1992} C. H. Bennett and S. J. Wiesner, Communication via one- and two-particle operators on Einstein-Podolsky-Rosen states,  Phys. Rev. Lett. {\bf 69}, 2881 (1992).

\bibitem{Bouw} D. Bouwmeester, J.-W. Pan, M. Mattle, M. Eible, H. Weinfurther, and A. Zeilinger, Experimental quantum teleportation,  Nature {\bf 390}, 575 (1997).

\bibitem{Bose} S. Bose, V. Vedral, and P. L. Knight, Multiparticle generalization of entanglement swapping,
Phys. Rev. A {\bf 57}, 822 (1998).

\bibitem{Barbieri} M. Barbieri, F. De Martini, G. Di Nepi, and P. Mataloni, Generation and Characterization of Werner States and Maximally Entangled Mixed States by a Universal Source of Entanglement,
Phys. Rev. Lett. {\bf 92}, 177901 (2004).

\bibitem{Joo} J. Joo, P. L. Knight, J. L. O'Brien, and T. Rudolph, One-way quantum computation with four-dimensional photonic qudits,  Phys. Rev. A {\bf 76}, 052326 (2007).

\bibitem{Stokes}  G. G. Stokes, On the Composition and Resolution of Streams of Polarized Light from different Sources,  Trans. Cambridge Philos. Soc. {\bf 9}, 399 (1852).

\bibitem{Mandel} L. Mandel and E. Wolf, {\it Optical Coherence and Quantum Optics}, Cambridge University
Press, Cambridge, UK, 1995. See pp. 342-355.

\bibitem{Fano-49} U. Fano, Remarks on the classical and quantum-mechanical treatment of partial polarization,
J. Opt. Soc. Am. {\bf 39}, 859 (1949).

\bibitem{Collet-70} E. Collet, Stokes Parameters for Quantum Systems, Am. J. Phys. {\bf 38}, 563 (1970).

\bibitem{Simon} R. Simon, Nondepolarizing systems and degree of polarization,  Opt. Commun. {\bf 77}, 349 (1990).

\bibitem{Kl-92} D. N. Klyshko, Multiphoton interference and polarization effects,
Phys. Lett. A {\bf 163}, 349 (1992).

\bibitem{Alod} A. P. Alodjants and S. M. Arakelian, Quantum phase measurements and non-classical polarization states of light,  J. Mod. Opt. {\bf 46}, 475 (1999).

\bibitem{Klimov-2010} A. B. Klimov, G. Bj\"{o}rk, J. S\"{o}derholm, L. S. Madsen, M. Lassen, U. L. Andersen,
J. Heersink, R. Dong, Ch. Marquardt, G. Leuchs, and L. L. S\'anchez-Soto, Assessing the Polarization of a Quantum Field from Stokes Fluctuations, Phys. Rev. Lett. {\bf 105}, 153602 (2010).

\bibitem{Bjork-2012} G. Bj\"{o}rk, J. S\"{o}derholm, Y.-S. Kim, Y.-S. Ra, H.-T. Lim, C. Kothe, Y.-H. Kim,
L. L. S\'anchez-Soto, A. B. Klimov, Central-moment description of polarization for quantum states of light,
Phys. Rev. A {\bf 85}, 053835 (2012).

\bibitem{Hoz-2013} P. de la Hoz, A. B. Klimov, G. Bj\"{o}rk, Y-H Kim, C. Muller, C. Marquardt, G. Leuchs,
and L. L. S\'anchez-Soto, Multipolar hierarchy of efficient quantum polarization measures,
Phys. Rev. A {\bf 88}, 063803 (2013).

\bibitem{Sanchez-2013} L. L. S\'anchez-Soto, A. B. Klimov, P. de la Hoz, and G. Leuchs, Quantum versus classical polarization states: when multipoles count, J. Phys. B {\bf 46}, 104011 (2013).

\bibitem{Hoz-2014} P. de la Hoz, G. Bj\"{o}rk, A. B. Klimov, G. Leuchs, and L. L. S\'anchez-Soto, Unpolarized states and hidden polarization, Phys. Rev. A {\bf 90}, 043826 (2014).

\bibitem{Bjork-2015} G. Bj\"{o}rk, A. B. Klimov, P. de la Hoz, M. Grassl, G. Leuchs, L. L. S\'anchez-Soto,
Extremal quantum states and their Majorana constellations,  Phys. Rev. A {\bf 92}, 031801(R) (2015).

\bibitem{Bjork-phys-scr-2015} G. Bj\"{o}rk, M. Grassl, P. de la Hoz, G. Leuchs, L. L. S\'anchez-Soto,
Stars of the quantum Universe: extremal constellations on the Poincar\'{e} sphere,
Phys. Scripta {\bf 90}, 108008 (2015).

\bibitem{Filip-2016} L. Ruppert, V. C. Usenko, and R. Filip, Estimation of the covariance matrix of macroscopic quantum states,  Phys. Rev. A {\bf 93}, 052114 (2016).

\bibitem{Klimov-2005} A. B. Klimov, L. L. S\'anchez-Soto, E. C. Yustas, J. S\"{o}derholm, and
G. Bj\"{o}rk, Distance-based degrees of polarization for a quantum field,
Phys. Rev. A {\bf 72}, 033813 (2005).

\bibitem{Bjork-2006} L. L. S\'anchez-Soto, J. S\"{o}derholm, E. C. Yustas, A. B. Klimov, and G. Bj\"{o}rk,
Degrees of polarization for a quantum field,  J. Phys.: Conf. Ser. {\bf 36}, 177 (2006).

\bibitem{Ghiu-2010} I. Ghiu, G. Bj\"{o}rk, P. Marian, and T. A. Marian, Probing light polarization with the quantum Chernoff bound, Phys. Rev. A {\bf 82}, 023803 (2010).

\bibitem{Bjork-2010} G. Bj\"{o}rk, J. S\"{o}derholm, L. L. S\'anchez-Soto, A. B. Klimov, I. Ghiu,
P. Marian, and T. A. Marian, Quantum degrees of polarization, Opt. Comm. {\bf 283}, 4440 (2010).

\bibitem{Ghiu-rom-ac-2015} I. Ghiu, C. Ghiu, and A. Isar, Quantum Chernoff degree of polarization of the Werner state, Proc. Rom. Acad. A {\bf 16}, 499 (2015).

\bibitem{Ghiu-rom-j-2016} I. Ghiu and A. Isar, The analytical expression of the Chernoff polarization of the Werner state, Rom. J. Phys. {\bf 61}, 768 (2016).

\bibitem{Ghiu-rom-rep-2018} I. Ghiu, Entanglement versus quantum degree of polarization, Rom. Rep. Phys. {\bf 70}, 104 (2018).

\bibitem{Chirkin} A. S. Chirkin, Polarization-squeezed light and quantum degree of polarization,
Optics and Spectroscopy {\bf 119}, 371 (2015).

\bibitem{Luis-2016} A. Luis, Polarization in quantum optics, Progress in Optics {\bf 61}, 283 (2016).

\bibitem{GJ} A. Z. Goldberg and D. F. V. James, Perfect polarization for arbitrary light beams,  Phys. Rev. A {\bf 96}, 053859 (2017).

\bibitem{Korolkova} N. Korolkova, G. Leuchs, R. Loudon, T. C. Ralph, and C. Silberhorn,
Polarization squeezing and continuous-variable polarization entanglement,
 Phys. Rev. A {\bf 65}, 052306 (2002).

\bibitem{Bowen} W. P. Bowen, R. Schnabel, H. -A. Bachor, P. K. Lam,
Polarization Squeezing of Continuous Variable Stokes Parameters,
Phys. Rev. Lett. {\bf 88}, 093601 (2002).

\bibitem{Luis-2007} A. Luis,  Polarization distributions and degree of polarization for quantum Gaussian light fields,  Opt. Comm. {\bf 273}, 173 (2007).

\bibitem{Muller-2016} C. R. M\"{u}ller, L. S. Madsen, A. B. Klimov, L. L. S\'anchez-Soto, G. Leuchs,
C. Marquardt, and U. L. Andersen, Parsing polarization squeezing into Fock layers,
Phys. Rev. A {\bf 93}, 033816 (2016).

\bibitem{Singh} R. S. Singh and H. Prakash, On the polarization of non-Gaussian optical quantum field: Higher-order optical polarization,  Ann. Phys. {\bf 333}, 198 (2013).

\bibitem{Opa} T. Opatrn\'y, G. Kurizki, and D.-G. Welsch, Continuous-variable teleportation improvement by photon subtraction via conditional measurement,  Phys. Rev. A {\bf 61}, 032302 (2000).

\bibitem{Olivares} S. Olivares, M. G. A. Paris, and R. Bonifacio, Teleportation improvement by inconclusive photon subtraction,  Phys. Rev. A {\bf 67}, 032314 (2003).

\bibitem{Anno} F. Dell'Anno, S. De Siena, L. Albano, and F. Illuminati, Continuous-variable quantum teleportation with non-Gaussian resources,  Phys. Rev. A {\bf 76}, 022301 (2007).

\bibitem{Eisert} J. Eisert, S. Scheel, and M. B. Plenio, Distilling Gaussian States with Gaussian Operations is Impossible,  Phys. Rev. Lett. {\bf 89}, 137903 (2002).

\bibitem{Cerf} N. J. Cerf, O. Kr\"{u}ger, P. Navez, R. F. Werner, and M. M. Wolf, Non-Gaussian Cloning of Quantum Coherent States is Optimal,  Phys. Rev. Lett. {\bf 95}, 070501 (2005).

\bibitem{Prakash} H. Prakash and N. Chandra, Density Operator of Unpolarized Radiation, Phys. Rev. A {\bf 4}, 796 (1971); {\em ibid.}
Phys. Rev. A {\bf 9}, 1021 (1974).

\bibitem{Bjork5} J. S\"{o}derholm, G. Bj\"{o}rk, and A. Trifonov, 	
Unpolarized Light in Quantum Optics,
Optics and Spectroscopy {\bf 91}, 532 (2001).

\bibitem{Jozsa} R. Jozsa, Fidelity for mixed quantum states, J. Mod. Opt. {\bf 41}, 2315 (1994).

\bibitem{Ved} V. Vedral, The role of relative entropy in quantum information theory,  Revs. Mod. Phys. {\bf 74}, 197 (2002).

\bibitem{Weh} A. Wehrl, General properties of entropy,  Revs. Mod. Phys. {\bf 50}, 221 (1978).

\bibitem{VP} V. Vedral and M. B. Plenio, Entanglement measures and purification procedures, Phys. Rev. A {\bf 57}, 1619 (1998).

\bibitem{PT2013} P. Marian and T.A. Marian, Relative entropy is an exact measure of non-Gaussianity,
Phys. Rev. A {\bf 88}, 012322 (2013).

\bibitem{AT}G. S. Agarwal and K. Tara,  Nonclassical character of states exhibiting no squeezing or sub-Poissonian statistics, Phys. Rev. A {\bf 46},  485 (1992).

\bibitem{bel}A.  Zavatta,  V. Parigi, and  M. Bellini, Experimental non-classicality of single-photon-added thermal light states,  Phys. Rev. A {\bf 75},  052106 (2007).

\bibitem{bel1}
T. Kiesel,  W. Vogel,  V. Parigi,  A. Zavatta, and  M. Bellini, Experimental determination of a nonclassical Glauber-Sudarshan $P$ function Phys. Rev. A {\bf 78},  021804R (2008).

\bibitem{Kis}T. Kiesel, W. Vogel, M. Bellini, and A. Zavatta, Nonclassicality quasiprobability of single-photon-added thermal states,  Phys. Rev. A {\bf 83},  032116 (2011).

\bibitem{Marian-2013} P. Marian, I. Ghiu, and T. A. Marian, Gaussification through decoherence,
Phys. Rev. A {\bf 88}, 012316 (2013).

\bibitem{Marian-ps-2013} I. Ghiu, P. Marian, and T. A. Marian, Measures of non-Gaussianity for
one-mode field states,  Phys. Scripta {\bf T153}, 014028 (2013).







\end{thebibliography}
\end{document}